\documentclass[notitlepage,english,aps,floats,onecolumn,10pt,showpacs,nofootinbib,floatfix]{revtex4-1}

\usepackage{pslatex}
\usepackage[T1]{fontenc}
\usepackage[latin1]{inputenc}
\usepackage{graphicx}
\usepackage{epsfig}
\usepackage{longtable}
\usepackage{calc}
\usepackage{ifthen}
\usepackage{amsmath}
\usepackage{hyperref}
\usepackage{amssymb}

\usepackage{color}

\newcommand{\nc}{\newcommand}
\nc{\renc}{\renewcommand}
\nc{\eqs}[2]{\mbox{Eqs.~(\ref{#1},\,\ref{#2})}}
\nc{\eq}[1]{\mbox{Eq.~(\ref{#1})}}
\nc{\figs}[2]{\mbox{Figs.~(\ref{#1},\,\ref{#2})}}
\nc{\fig}[1]{\mbox{Fig~.(\ref{#1})}}
\nc{\be}[1]{\begin{equation} \mbox{$\label{#1}$}}
\nc{\ee}{\vspace{0.1cm}\end{equation}}

\newcommand{\bean}{\begin{eqnarray*}}
\newcommand{\eean}{\end{eqnarray*}}

\def\GeV{{\rm \ GeV}}

\def\lae{\;^{<}_{\sim} \;} \def\gae{\; ^{>}_{\sim} \;}

\begin{document}

\title{Sub-Planckian $\phi^{2}$ Inflation in the Palatini Formulation of Gravity with an $R^2$ term} 

\author{Amy Lloyd-Stubbs  and John McDonald }
\email{a.lloyd-stubbs@lancaster.ac.uk}
\email{j.mcdonald@lancaster.ac.uk}
\affiliation{Dept. of Physics,  
Lancaster University, Lancaster LA1 4YB, UK}

\begin{abstract}

The simplest model that can produce inflation is a massive non-interacting scalar particle with potential $V = m^{2} \phi^{2}/2$. However, $\phi^{2}$ chaotic inflation is inconsistent with the observed upper bound on the tensor-to-scalar ratio, $r$. Recently it has been shown that, in the context of the Palatini formalism of gravity with an $R^{2}$ term, the $\phi^{2}$ potential can be consistent with the observed bound on $r$ whilst retaining the successful prediction for the scalar spectral index, $n_{s}$. Here we show that the Palatini $\phi^{2} R^2$ inflation model can also solve the super-Planckian inflaton problem of $\phi^{2}$ chaotic inflation, and that the model can be consistent with Planck scale-suppressed potential corrections, as may arise from a complete quantum gravity theory. If $\alpha \gae 10^{12}$, where $\alpha$ is the coefficient of the $R^2$ term,  the inflaton in the Einstein frame, $\sigma$, remains sub-Planckian throughout inflation. In addition, if $\alpha \gae 10^{20}$ then the predictions of the model are unaffected by Planck-suppressed potential corrections in the case where there is a broken shift symmetry,  and if
$\alpha \gae 10^{32}$ then the predictions are unaffected by Planck-suppressed potential corrections in general. The value of $r$ is generally small, with $r \lae 10^{-5}$ for $\alpha \gae 10^{12}$. We calculate the maximum possible reheating temperature, $T_{R\;max}$, corresponding to instantaneous reheating, for the different regimes of $\alpha$. We find that for $\alpha \approx 10^{32}$, $T_{R\; max}$ is approximately $10^{10}$ GeV, with larger values of $T_{R\;max}$ for smaller $\alpha$. For the case of instantaneous reheating, we show that $n_{s}$  is in agreement with the 2018 Planck results to within 1-$\sigma$, with the exception of the $\alpha \approx 10^{32}$ case, which is close to the 2-$\sigma$ lower bound. Following inflation, the inflaton condensate is likely to rapidly fragment, which makes it possible for reheating to occur via the Higgs portal due to inflaton annihilations within oscillons. This typically results in delayed reheating, which is disfavoured by the observed value of $n_{s}$. In contrast, reheating via inflaton decays to right-handed neutrinos can easily result in instantaneous reheating. We determine the scale of unitarity violation and show that, in general, unitarity is conserved during inflation, although the inflaton field is larger than the unitarity-violation scale. We conclude that the Palatini $\phi^{2} R^{2}$ inflation model provides a completely consistent model of inflation which can be sub-Planckian and consistent with Planck scale-suppressed potential corrections, can reheat successfully, and conserves unitarity during inflation.

\end{abstract}
 
\maketitle

\section{Introduction}

One of the simplest models of inflation is $\phi^{2}$ chaotic inflation. However, although the prediction of the model for the scalar spectral index, $n_s$, is in excellent agreement with observations, the model has been ruled out observationally due to its large prediction for the tensor-to-scalar ratio, $r$. Nevertheless, the possibility of using a simple renormalisable potential which can easily connect to particle physics models is very attractive from a model-building point of view. 
Recently, it has been shown in \cite{Enckell} and \cite{Antoniadis} that by considering a $\phi^2$ potential together with an $R^{2}$ term in the Palatini formalism\footnote{For a recent review of Palatini inflation models, see \cite{treview}.}\footnote{Natural inflation and quartic inflation have also been considered in the Palatini plus $R^2$ framework  \cite{karam}.}, it is possible to suppress the tensor-to-scalar ratio whilst preserving the successful prediction for the scalar spectral index.

In the standard metric formulation of gravity, the spacetime connection, $\Gamma$, specialises to the Levi-Civita connection, which depends on the spacetime metric, $g_{\mu \nu}$. In this case, the Ricci tensor and, by association, the Ricci scalar, both carry a dependence on the metric and derivatives of the metric. As an alternative, the Palatini formulation of gravity uses a form of the connection which does not depend on the spacetime metric \cite{palatini1}. Therefore the Ricci tensor and the Ricci scalar depend only the connection, $R_{\mu \nu}=R_{\mu \nu}\left(\Gamma \right)$ and $R=R\left(\Gamma \right)$. In a conventional General Relativity (GR) setting the two formalisms are equivalent, since the connection takes the Levi-Civita form once the equations of motion are applied; the difference arises in models where a non-minimal coupling of a scalar field to gravity or a higher-order term in $R$ is included. This is because in the metric formalism, when a conformal transformation is made to the Einstein frame, the Ricci tensor must also be transformed due to its dependence on the metric. Therefore this transformation leads to an additional kinetic term involving the conformal factor. In the Palatini case, because there is no metric dependence in the Ricci tensor, the transformation to the Einstein frame is much more straightforward, as the conformal factors only appear due to the transformation of the explicit metric in $R=g^{\mu \nu}R_{\mu \nu}$. This means that the results obtained in the metric and Palatini versions of an inflation model for the slow-roll parameters, scalar spectral index and tensor-to-scalar ratio are generally different \cite{palatini1}. 

In conventional $\phi^2$ chaotic inflation, the inflaton is super-Planckian, with $\phi \approx 15 M_{pl}$ at $N \approx 60$. Aside from general concerns over the consistency of a super-Planckian value of $\phi$ with theories which seek to unify with gravity, it is also possible that a complete quantum gravity theory will introduce Planck scale-suppressed operators into the potential, which can modify the predictions of the model at sub-Planckian inflaton values. (For a review, see \cite{baumann}.) Here we will investigate whether the Palatini $\phi^{2} R^2$ inflation model can also address the problems of a super-Planckian inflaton field and the consistency of the model with Planck scale-suppressed potential corrections. We will also consider whether the resulting models can serve as viable inflation models, with successful reheating and consistency with unitarity conservation during inflation.

The paper is organised as follows. In Section II we introduce the Palatini $\phi^{2} R^2$ model. In Section III we discuss the bounds on the dimensionless constant of the $R^2$ term in the action, $\alpha$, in order for the inflaton to be sub-Planckian and for $n_{s}$ to be consistent with Planck-scale suppressed potential corrections. In Section IV we discuss the reheating temperature under the assumption of instantaneous reheating and the resulting predictions for $n_{s}$. In Section V we discuss the condition for unitarity conservation during inflation. In Section VI we summarise our results for the case of instantaneous reheating. In Section VII we consider two specific mechanisms for reheating and the associated quantum corrections to the potential: inflaton decay to right-handed neutrinos and inflaton annihilation to Higgs bosons via the Higgs portal. In Section VIII we present our conclusions.

\section{$\phi^{2}$ Inflation in Palatini plus $R^2$ gravity}

 We consider the case of a $\phi^{2}$ potential in the limit where the non-minimal coupling of $\phi$ to the Ricci scalar $R$ is zero. In this limit the Jordan frame action of the model is\footnote{The Jordan frame may be thought of as the frame in which the model is {\it defined} i.e. in which the structure and symmetries of the model are apparent, whereas the Einstein frame is the frame in which physics and cosmology are conventional, corresponding to standard GR combined with minimally coupled, canonically normalised fields.}   \cite{Enckell, Antoniadis}  
\be{e1}  S = \int d^4 x \sqrt{-g}\left[ \frac{M_{pl}^2}{2}R + \frac{\alpha}{4} R^2 - \frac{1}{2} \partial_{\mu} \phi \partial^{\mu} \phi - V(\phi)  \right] ~,\ee
where the signature is $(-, +, +, +)$. The action \eq{e1} can be written in terms of an auxiliary field $\chi$
\be{e2}  S = \int d^4 x \sqrt{-g}\left[ \frac{1}{2}R (M_{pl}^{2} + \alpha \chi^2) - \frac{\alpha}{4} \chi^{4} - \frac{1}{2} \partial_{\mu} \phi \partial^{\mu} \phi - V(\phi)   \right] ~.\ee  
We generally consider inflation in the Einstein frame, which corresponds to conventional GR, with all transformed Einstein frame quantities denoted by a tilde for clarity. 
The Lagrangian is transformed to the Einstein frame via the conformal factor $\Omega$, where 
\be{e3}  \Omega^{2} = 1 + \frac{\alpha \chi^{2}}{M_{pl}^2}   ~.\ee
The conformal transformation in the Palatini formalism is  given by $\tilde{g}_{\mu\;\nu} = \Omega^{2} g_{\mu \; \nu}$ and $ \tilde{R} = R/\Omega^{2}$. The Einstein frame action is then
\be{e4} S_{E} = \int d^4 x \sqrt{-\tilde{g}}\left[ \frac{M_{pl}^{2}}{2} \tilde{R}  - \frac{\alpha \chi^{4}}{4 \Omega^4 } - \frac{1}{2 \Omega^{2}} \partial_{\mu} \phi \partial^{\mu} \phi - \frac{V(\phi)}{\Omega^{4}}   \right] ~.\ee  
On eliminating the auxiliary field $\chi$ via its equation of motion, the Einstein frame action becomes \cite{Enckell, Antoniadis}  
\be{e5}
S_{E}=\int d^{4}x \sqrt{-\tilde{g}} \left[\frac{1}{2}M_{pl}^{2}\tilde{R} -\frac{1}{2}\frac{\partial^{\mu}
\phi \partial_{\mu} \phi }{\left(1 + \frac{4\alpha V\left(\phi \right)}{M_{pl}^{4}} \right)} + \frac{ \alpha}{4 M_{pl}^{4}} 
\frac{\left(\partial_{\mu} \phi \partial^{\mu} \phi \right)^{2}}{\left(1 + \frac{4 \alpha V(\phi)}{M_{pl}^4}\right)}  
- \frac{V\left(\phi \right)}{\left(1 + \frac{4 \alpha V\left(\phi \right)}{M_{pl}^{4}}\right)}\right]
~.\ee
In the following we define the Jordan frame potential by   
\be{e8}
V\left(\phi \right) = \frac{m_{\phi}^{2} \phi^{2}}{2}
~. \ee
 The corresponding conformal factor is  
\be{e1b} \Omega^{2} \equiv  1 +  \frac{\alpha \left(4 V\left(\phi \right) + \partial_{\mu} \phi \partial^{\mu} \phi \right)}{M_{pl}^{4} - \alpha\partial_{\mu} \phi \partial^{\mu} \phi}  \approx  1 + \frac{2 \alpha m_{\phi}^{2} \phi^{2}}{M_{pl}^{4}} ~,\ee
where the latter expression is valid during slow-roll inflation, during which the derivative terms are negligible.

\subsection{Inflation in the Einstein frame}  

We will consider inflation in the Einstein frame, which is the appropriate frame for the analysis of inflaton dynamics and Planck-scale suppressed potential corrections. We first canonically normalise the kinetic term. In order to do this we define a canonically normalised scalar field $\sigma$, which is related to $\phi$ by
\be{e7}
\left(\frac{d\sigma}{d\phi}\right)^{2} = \frac{1}{1 + \frac{2 \alpha m_{\phi}^{2} \phi^{2}}{M_{pl}^{4}} }
\Rightarrow 
\frac{d\sigma}{d\phi} = \pm \frac{1}{\sqrt{ 1 + \frac{2 \alpha m_{\phi}^{2} \phi^{2}}{M_{pl}^{4}} } }
~.\ee
We will choose the positive solution in the following. 
Integrating this gives 
\be{e8}
\sigma = 
\frac{1}{\sqrt{K}} \ln \left( \sqrt{1 + K \phi^{2}} + \sqrt{K} \phi \right) + C \;\;\;;\;\; K = \frac{2 \alpha m_{\phi}^{2}}{M_{pl}^{4}} 
~,\ee
where $C$ is an integration constant. 
We will define $\sigma$ such that $\sigma \approx \phi$ when 
$\Omega \approx 1$, corresponding to $\phi < \phi_{0}$,  where
\be{e9}  \phi_{0} = \frac{M_{pl}^{2}}{\sqrt{2\alpha} m_{\phi} }  ~.\ee 
In this case 
\be{e10}
\sigma =  \phi_{0} \ln \left( \sqrt{1 + \frac{\phi^{2}}{\phi_{0}^{2}}} + \frac{\phi}{\phi_{0}}  \right)
~.\ee
Thus, to a good approximation, 
$$ \sigma \approx \phi \;\;\;;\;\; \phi < \phi_{0} \, ,  $$  
\be{e11} \sigma \approx \phi_{0} \ln\left(\frac{2 \phi}{\phi_{0}} \right) = \frac{M_{pl}^{2}}{\sqrt{2 \alpha} m_{\phi}} 
\ln \left(\frac{2 \sqrt{2 \alpha} m_{\phi} \phi}{M_{pl}^{2}} \right)  
\;\;\;;\;\; \phi > \phi_{0}  ~.\ee 
In the following we will derive $\sigma(N)$ and the inflation observables to leading order in $\phi_{0}^{2}/\phi^{2}$, which is very small during inflation in the models considered here. 
We define the Einstein frame potential by
\be{e11a}
V_{E}\left(\phi \right)=  \frac{V\left(\phi \right)}{1 + \frac{4\alpha V\left(\phi \right)}{M_{pl}^{4}}}  
~.\ee
To leading order in $\phi_{0}^{2}/\phi^{2}$ this becomes
\be{e12}
V_{E}\left(\phi \right) = \frac{M_{Pl}}{4 \alpha} \left(1 + \frac{\phi_{0}^{2}}{\phi^{2}} \right)^{-1}  \approx \frac{M_{pl}^{4}}{4\alpha}\left(1-\frac{M_{pl}^{4}}{2 \alpha m_{\phi}^{2} \phi^{2} } \right)
~.\ee
In terms of the canonically normalised field, the inflaton potential is therefore 
\be{e12a}  V_{E}(\sigma) \approx \frac{M_{pl}^{4}}{4 \alpha} \left( 1 - 4 \exp\left(\frac{- 2 \sqrt{2 \alpha} m_{\phi} \sigma}{M_{pl}^{2}}\right) \right)    ~.\ee
The number of e-folds of inflation in the Einstein frame is given by\footnote{In the Einstein frame action, the $\sigma$ derivative terms can be written as 
$$ -\frac{1}{2} \partial_{\mu} \sigma \partial^{\mu} \sigma + \frac{\alpha}{4} \left(1 + \frac{4 \alpha V(\phi)}{M_{pl}^{4}}\right) \frac{ \left( \partial_{\mu} \sigma \partial^{\mu} \sigma\right)^{2}}{M_{pl}^{4}}   \approx
-\frac{1}{2}\left(1 - \frac{2 \alpha^{2} V}{M_{pl}^{8}}  \partial_{\nu} \sigma \partial^{\nu} \sigma \right) \partial_{\mu} \sigma \partial^{\mu} \sigma  ~,$$
where in the latter expression we have assumed that $4  \alpha V(\phi)/M_{pl}^{4} \gg 1$ during inflation. 
Substituting the slow-roll expression for $\sigma(t)$, we find that    
$$  1 - \frac{2 \alpha^{2} V}{M_{pl}^{8}}  \partial_{\nu} \sigma \partial^{\nu} \sigma = 1 + \frac{1}{24 N}  ~.$$   
Therefore the quartic derivative term is negligible during slow-roll inflation and the conventional analysis of slow-roll inflation applies.}   
\be{e13}
N\left(\sigma \right) = - \frac{1}{M_{pl}^{2}} \int_{\sigma}^{\sigma_{end}} \frac{V_{E}}{V^{'}_{E}} d \sigma  
\approx \frac{M_{pl}^{2}}{32 \alpha m_{\phi}^{2}} 
\exp{\left(\frac{2 \sqrt{2 \alpha} m_{\phi} \sigma}{M_{pl}^{2}}\right)}
~,\ee
where $V_{E} \approx M_{pl}^{4}/4\alpha$ (since $\phi_{0}^{2}\ll \phi^{2}$ during inflation) and we have assumed that $\sigma_{end} <<  \sigma$, where $\sigma_{end}$ is the value of $\sigma$ at the end of slow-roll inflation. Therefore $\sigma(N)$ is given by  
\be{e14}
\sigma \left(N\right) \approx \frac{M_{pl}^{2}}{2 \sqrt{2 \alpha} m_{\phi} }
\ln\left(\frac{32 \alpha m_{\phi}^{2} N }{M_{pl}^{2}} \right)  
~.\ee 
\eq{e11} then implies that 
\be{e19a}  \phi(N) \approx 2 \sqrt{N} M_{pl}   ~\ee
On substituting $\sigma(N)$ into the $\eta$ and $\epsilon$ parameters in the Einstein frame, the leading-order slow-roll parameters and inflation observables are given by  
\be{e15}
\epsilon = \frac{M_{pl}^{2}}{2} \left(
\frac{ \frac{\partial V_{E}}{\partial \sigma} }{V_{E}}
\right)^{2} 
\approx \left(\frac{64 \alpha m_{\phi}^{2}}{M_{pl}^{2}} \right) \exp\left(- \frac{4 \sqrt{2 \alpha } m_{\phi} \sigma}{M_{pl}^{2}}  \right)
\Rightarrow \epsilon \approx 
\frac{M_{pl}^{2}}{16 \alpha m_{\phi}^{2} } 
 \frac{1}{N^{2}}  
~,\ee
\be{e16}
\eta =  M_{pl}^{2} \frac{ \frac{\partial^{2} V_{E}}{\partial \sigma^{2}} }{V_{E}} \approx - \left(\frac{32 \alpha m_{\phi}^{2}}{ M_{pl}^{2}} \right) 
\exp\left( - \frac{ 2 \sqrt{2 \alpha} m_{\phi} \sigma}{M_{pl}^{2}} 
\right) \Rightarrow \eta \approx 
 -\frac{1}{N}
~,\ee
\be{e17}
n_{s} = 1 + 2 \eta - 6 \epsilon \approx 1 - \frac{2}{N} 
~,\ee
\be{e18}
r = 16 \epsilon  \approx \frac{M_{pl}^{2}}{\alpha m_{\phi}^{2}} 
\frac{1}{N^{2}} ~,\ee
and
\be{e19}
\alpha_{s} = - \frac{d n_{s}}{dN} \approx -\frac{2}{N^{2}} 
~.\ee
Note that 
\be{e19a} \epsilon \equiv \frac{1}{2N} \frac{\phi_{0}^{2}} {\phi^{2}(N)}   ~,\ee
therefore very small $\epsilon$ corresponds to very small $\phi_{0}^{2}/\phi^{2}$.  
Our Einstein frame results for $n_{s}$, $r$ and $\alpha_{s}$ are in complete agreement with those of the general analysis given in \cite{Enckell}.

\subsection{End of Slow-Roll Inflation} 

$\sigma_{end}$ is defined by $|\eta(\sigma)| \approx 1$. Assuming that slow-roll inflation ends when $\sigma$ is on the plateau of the potential, we find that
\be{e22}
\sigma_{end} \approx
\frac{M_{pl}^{2}}{2 \sqrt{2 \alpha} m_{\phi} }   
\ln\left( \frac{32 \alpha m_{\phi}^{2}}{M_{pl}^{2}} \right) ~.\ee
The corresponding value of $\phi$ at the end of inflation is 
\be{e23}
\phi_{end}  \approx \frac{M_{pl}^{2}}{2 \sqrt{2 \alpha} m_{\phi} } 
\exp{\left(\frac{\sqrt{2 \alpha} m_{\phi} \sigma_{end}}{M_{pl}^{2} } \right)} = 2 M_{pl}   ~.\ee

\subsection{Power spectrum and $m_{\phi}$} 

On substituting our expression for $\epsilon \left(N\right)$ and
$V_{E} \approx M_{pl}^{4}/4\alpha$ 
into the standard expression for the power spectrum, we obtain
\be{e20}
P_{\textit{R}} \equiv \frac{V_{E}}{24\pi^{2}\epsilon M_{pl}^{4}}
= \frac{m_{\phi}^{2} N^{2}}{6 \pi^{2} M_{pl}^{2}}   
\Rightarrow 
m_{\phi} = \frac{\sqrt{6} \pi M_{pl} P_{\textit{R}}^{1/2} }{N}  
~.\ee
To find $m_{\phi}$, we use $N=60$ as an estimate for the Planck pivot scale for now, and the observed value of the power spectrum from Planck, $P_{\textit{R}}= 2.1 \times 10^{-9}$, which gives   
\be{e27}
m_{\phi} = 1.4 \times 10^{13} \GeV  ~.\ee

\section{Sub-Planckian $\phi^{2}$ Inflation and Planck-suppressed potential corrections} 

In conventional $\phi^2$ chaotic inflation, the inflaton field is greater than the Planck scale during inflation, with $\phi \approx 15 M_{pl}$ at $N \approx 60$. Beyond the question of super-Planckian field values in theories which seek to unify gravity with other forces - which suggest that such field values cannot be achieved \cite{baumann} - there is also the question of how corrections associated with a UV completion of quantum gravity will affect inflation observables. In the following we will determine the constraints on the model from: (i) the requirement of a sub-Planckian inflaton during inflation, (ii) the effect of general Planck-scale suppressed potential corrections on inflation observables, and (iii) the effect of Planck-scale suppressed potential corrections in the case of a broken shift symmetry.

In the present model, the Planck energy is the energy at which quantum gravity fails in the Einstein frame, since this is the frame in which conventional GR applies and in which unitarity is violated at the Planck energy by graviton scattering. Therefore the Planck scale should be interpreted as the scale of the UV completion of gravity in the Einstein frame.  In the following we will consider the sub-Planckian requirement and the Planck-scale potential corrections to apply in the Einstein frame.

In the case where the Planck scale is the cut-off scale of the effective theory of the UV completion of quantum gravity, all possible non-renormalisable operators which are consistent with the symmetries of the UV completion are expected to appear in the low-energy effective theory. Therefore all possible Planck-suppressed operators for the canonically normalised scalar $\sigma$ should be added to the Einstein frame Lagrangian. We therefore consider non-renormalisable potential terms of the form 
\be{e31}
\Delta V_{E} = \sum_{n} \frac{k_{n} \sigma^{n}}{M_{pl}^{n-4}} 
~,\ee
where dimensionally we expect $k_{n} \sim 1$, and we will assume a symmetry $\sigma \leftrightarrow -\sigma$ of the non-renormalisable terms, consistent with the $\phi \leftrightarrow -\phi$ symmetry of the $\phi^{2}$ potential\footnote{\eq{e7} is invariant under $\phi \leftrightarrow -\phi$ and $\sigma \leftrightarrow -\sigma$. Therefore if the Jordan frame action is invariant under $\phi \leftrightarrow -\phi$, the transformed action will be invariant under $\sigma \leftrightarrow -\sigma$. This means that $\sigma$ in the Einstein frame potential \eq{e12a} should be replaced by $|\sigma|$ when $\sigma < 0$.}.   In the following we will focus on the leading-order potential correction, corresponding to $n = 6$ 
\be{e32}  \Delta V_{E} =   \frac{k \sigma^{6}}{M_{pl}^{2}} ~,\ee
where we have written $k_{6}$ as $k$. \eq{e32} is expected if there are no further symmetries of the complete theory to forbid it. In the case of conventional $\phi^2$ chaotic inflation, it has been proposed that there could be a shift symmetry of the complete theory, $\phi \rightarrow \phi + constant$, which is broken by the mass squared term in the renormalisable potential. In this case, any non-renormalisable corrections to the potential should vanish as $m_{\phi}^2 \rightarrow 0$ and therefore should be proportional to $m_{\phi}^2$. The same assumption can be applied to the Palatini $\phi^2 R^2$ model, since the Einstein frame potential vanishes as $m_{\phi}^2 \rightarrow 0$. Under the assumption that $m_{\phi}^{2}$ is the shift symmetry-breaking parameter, the leading-order non-renormalisable term in the potential in the Einstein frame has the form 
\be{e33} \Delta V_{E} \approx   \frac{m_{\phi}^2 \sigma^{6}}{M_{pl}^{4}} ~.\ee
This term will have a weaker effect on the inflation observables. We will consider both possibilities \eq{e32} and \eq{e33} in the following.

\subsection{Bound on $\alpha$ from sub-Planckian $\sigma$ during inflation} 

We first derive the constraint on $\alpha$ by imposing that $\sigma$ remains sub-Planckian during inflation
\be{e46}
\sigma\left(N\right) < M_{pl} ~.
\ee
Substituting the expression for $\sigma\left(N\right)$, \eq{e14},  we obtain the constraint
\be{e47} \sqrt{\frac{2 \alpha m_{\phi}^{2}}{M_{pl}^{2}} } > \frac{1}{2}\ln\left(\frac{32 \alpha m_{\phi}^{2} N}{ M_{pl}^{2}}\right)~.
\ee
Using this constraint we find that, in order to keep $\sigma$ sub-Planckian, $\alpha$ must satisfy  
\be{e48} \alpha \gae 10^{12} ~.\ee

While we should consider the sub-Planckian condition in the Einstein frame, we note that the condition \eq{e48} can also be broadly understood in the Jordan frame as the condition for $\phi$ during inflation to be less than the effective Planck mass $M_{pl,\,eff} \equiv \Omega M_{pl}$, where the conformal factor is given by \eq{e1b}. During inflation, the effective Planck mass is 
\be{cf1} M_{pl,\,eff}^{2} = \Omega^{2} M_{pl}^{2} \approx 
\frac{2 \alpha m_{\phi}^{2} \phi^{2}}{M_{pl}^{2}}  ~.\ee
The condition that $\phi < M_{pl\;eff}$ during inflation is therefore 
\be{cf2}  \alpha >  \frac{M_{pl}^{2}}{2 m_{\phi}^{2}}  = 1.5 \times 10^{10} \GeV ~.\ee
This condition is satisfied whenever the Einstein frame condition \eq{e48} is satisfied. However, it is significantly weaker, showing that the sub-Planckian condition needs to be considered in the Einstein frame, where the Planck mass and its relation to gravity is well-defined.

\subsection{Bound on $\alpha$ from Planck-scale suppressed potential corrections} 

We next derive a lower bound on $\alpha$ from the shift of the scalar spectral index due to the leading-order Planck-suppressed potential correction, \eq{e32}. 
In this case, the Einstein frame potential takes the form
\be{e38}
V_{TOT} \equiv V_{E} + \Delta V_{E} = V_{E}\left(\sigma \right) + \frac{k \sigma^{6}}{M_{pl}^{2}} 
~. \ee
Since $\epsilon << 1$ in the model of interest, the scalar spectral index is approximately
\be{e39}
n_{s} \thickapprox 1 + 2\eta
~,\ee 
where 
\be{e40}
\eta = M_{pl}^{2}\frac{V_{TOT}''}{V_{TOT}} \approx M_{pl}^{2}\frac{V_{E}'' + \Delta V_{E}''}{V_{E}}  ~,
\ee
and where $\Delta V_{E} \ll V_{E}$ such that $V_{TOT} \approx V_{E}$. 
The $\eta$ shift is then given by
\be{e41}
\Delta \eta \approx M_{pl}^{2} \frac{\Delta V_{E}''}{V_{E}}
\Rightarrow \Delta \eta \approx \frac{120k\alpha}{M_{pl}^{4}}\sigma^{4}.
\ee
Substituting $\sigma(N)$ into this expression,  we obtain
\be{e43}
\Delta \eta \approx \frac{30 M_{pl}^{4} k}{m_{\phi}^{4} \alpha} 
\ln^{4}\left(\frac{4 \sqrt{2 \alpha} m_{\phi} \sqrt{N}}{M_{pl}} \right)
~.\ee
In order to preserve the successful prediction of $n_{s}$ we impose the constraint
\be{e44}
\mid \Delta \eta \mid < 0.001   ~.\ee
Using the value of $m_{\phi}$ obtained earlier and $k = 1$, we find that constraint \eq{e44} imposes the lower bound    
\be{e45} \alpha \gae 1.5 \times 10^{31}   ~.\ee 

In this it is assumed that the expression for $\sigma$ as a 
function of $N$, \eq{e14}, is unaffected by the potential correction, 
which is essential for the successful prediction of $n_{s}$. 
This requires that the contribution of $\Delta V'_{E}$ to the 
$\sigma$ field equation is small compared to that of $V'_{E}$. This can be stated more precisely by considering the expression for $N$  
\be{e45x} N = - \frac{1}{M_{pl}^{2}}
\int_{\sigma}^{\sigma_{end}} \frac{V_{TOT}}{V'_{TOT}} d \sigma 
\approx   - \frac{1}{M_{pl}^{2}} \int_{\sigma}^{\sigma_{end}} \frac{V_{E}}{V'_{E} \left(1 + \frac{\Delta V_{E}'}{V_{E}'} \right)} d \sigma 
\approx - \frac{1}{M_{pl}^{2}} \int_{\sigma}^{\sigma_{end}} \frac{V_{E}}{V'_{E}} \left(1 - \frac{\Delta V_{E}'}{V_{E}'} \right)  d \sigma  ~.\ee
Therefore $|\Delta V_{E}'/V_{E}'| \approx 0.1$ will change $N$ by $|\Delta N/N| \sim 0.1$ and so $|\Delta \eta| = |\Delta (1/N)| = |\Delta N|/N^{2} \sim 0.001$. Thus $|\Delta \eta| \lae 0.001$ requires that $|\Delta V_{E}'(\sigma)| \lae  0.1 |V_{E}^{'}(\sigma)| $. $\Delta V'_{E}/V'_{E}$ is given by
\be{e45y}  \frac{ \Delta V'_{E}}{V'_{E} } = \frac{ 3 k \sqrt{\alpha}}{\sqrt{2} m_{\phi} M_{pl}^{4}} \sigma^{5} \exp\left( \frac{2 \sqrt{2 \alpha} m_{\phi} \sigma}{M_{Pl}^{2}} \right) ~.\ee
Using $\sigma(N)$ from \eq{e14}, this becomes 
\be{e45z} \frac{ \Delta V'_{E}}{V'_{E} } = \frac{3 k N}{8 \alpha} \frac{M_{pl}^{4}}{m_{\phi}^{4}} \ln^{5} \left(\frac{32 \alpha m_{\phi}^{2} N}{M_{pl}^{2} } \right)   ~.\ee
Therefore  $|\Delta V_{E}'(\sigma)| \lae  0.1 |V_{E}^{'}(\sigma)| $ is satisfied if
\be{e45b}  \alpha \gae 10 \times \frac{3 k N}{8} \frac{M_{pl}^{4}}{m_{\phi}^{4}} \ln^{5} \left(\frac{32 \alpha m_{\phi}^{2} N}{M_{pl}^{2} } \right)   ~.\ee
With $N = 60$ and $k = 1$, this requires that 
\be{e45c}  \alpha \gae 1.2 \times 10^{32}  ~.\ee
Therefore the condition that the scalar spectral index is not significantly changed by Planck-suppressed potential corrections requires that $\alpha \gae 10^{32}$.     

\subsection{Bound on $\alpha$ from Planck-scale potential corrections with a shift symmetry} 

The leading-order Planck-suppressed corrections in the Einstein frame in the case of a shift symmetry is
\be{e50} \Delta V_{E} \approx \left(\frac{m_{\phi}^{2}}{M_{pl}^{2}}\right)\frac{\sigma^{6}}{M_{pl}^{2}}
~.\ee
The effective value of $k$ in \eq{e43} is then modified from $k \sim 1$ to 
\be{e51}
k \sim \frac{m_{\phi}^{2}}{M_{pl}^{2}}
~.\ee
We follow the same treatment as before to calculate the lower bound on $\alpha$ needed to suppress the shift of the scalar spectral index. In this case we find that the condition that $n_{s}$ is not significantly changed becomes
\be{e52}
\alpha \gae 3.6 \times 10^{19} 
~.\ee
Therefore the scalar spectral index will remain in agreement with Planck if $\alpha \gae 10^{20}$. 

\section{Instantaneous Reheating Temperature $T_{R\;max}$ and  $n_{s}$}

In order to accurately determine the prediction for $n_{s}$ at the Planck pivot scale, we need to know the corresponding value of $N$. This will depend on the value of the expansion rate in the Einstein frame during inflation and the reheating temperature. 

We first show that inflation ends and rapid oscillations of $\sigma$  begin when the inflaton is clearly on the plateau of the potential.
The value of the inflaton field at which the inflaton transitions from the plateau to a $\sigma^{2}$ potential is $\sigma_{0}$, where
\be{e54}
\sigma_{0} \approx \phi_{0} = \frac{M_{pl}^{2}}{\sqrt{2 \alpha} m_{\phi}} 
~.\ee
Therefore 
\be{e55}
\frac{\sigma_{end}}{\sigma_{0}} \approx \frac{1}{2}\ln\left(\frac{32 \alpha m_{\phi}^{2}}{M_{pl}^{2}} \right)  ~.\ee
The value of $32 \alpha m_{\phi}^{2}/M_{pl}^{2}$ ranges from 1100 to $1.1 \times 10^{23}$ for $\alpha$ from $10^{12}$ to $10^{32}$  (corresponding to the range of lower bounds on $\alpha$ from the sub-Planckian limit to the limit from generally suppressed Planck potential corrections). This implies that $\sigma_{end}/\sigma_{0}$ ranges from 3.5 to 26.5. Therefore inflation ends when $\sigma$ is clearly on the plateau, $\sigma_{end} > \sigma_{0}$. The Hubble parameter at the end of inflation in the Einstein frame, $\tilde{H}$, is therefore the same as $\tilde{H}$ during inflation, 
\be{e56}
\tilde{H} \approx \left(\frac{V_{E}}{3M_{pl}^{2}}\right)^{\frac{1}{2}}
\approx \frac{M_{pl}}{\sqrt{12\alpha}}
~,\ee
where $V_{E} \approx M_{pl}^{4}/4 \alpha$ on the plateau.

The energy density during inflation converts to rapid oscillations of the field once slow roll inflation ends at 
$\sigma_{end}$. The assumption that the energy density during inflation, $\tilde{\rho} = V_{E}$, instantly decays to radiation then gives the maximum possible reheating temperature, $T_{R\;max}$, which is related to $\tilde{\rho}$ by 
\be{e58}
\tilde{\rho} \equiv 3M_{pl}^{2} \tilde{H}^{2} = 
\frac{\pi^{2}}{30} g\left(T_{R\;max}\right)T_{R\;max}^{4}
~.\ee
Therefore
\be{e60}
T_{R\;max}
 = 
\left( \frac{15}{2 \pi^{2} g(T_{R\;max})}\right)^{1/4} \frac{M_{pl}}{\alpha^{1/4}}  
~.\ee
Assuming instantaneous reheating at the end of inflation and a constant value of $\tilde{H}$ during inflation, the number of e-foldings of inflation at which a present length scale $\lambda_{0}$ exits the horizon is
\be{e58}
N = \ln{\left[\left(\frac{g_{s}\left(T_{0}\right)}{g_{s}\left(T_{R\;max}\right)}\right)^{\frac{1}{3}}\frac{T_{0}\lambda_{0} \tilde{H}}{T_{R\;max}} \right]}
~,\ee
where $T_{0}$ is the present Cosmic Microwave Background (CMB) temperature. Therefore the number of e-folds $N$ corresponding to $\lambda_{0}$ is
\be{e62}
N = \ln \left[{\left(\frac{g_{s}\left(T_{0}\right)}{g_{s}\left(T_{R\;max}\right)}\right)^{\frac{1}{3}}\left(\frac{\pi^{2}g\left(T_{R\;max}\right)}{1080\alpha}\right)^{\frac{1}{4}}T_{0} \lambda_{0}} \right]
~.\ee
The Planck pivot scale corresponds to $k \equiv 2\pi/\lambda_{0} =   0.05 \, {\rm Mpc^{-1}} \equiv  3.2 \times 10^{-40}$ GeV, therefore $\lambda_{0} = 2\pi/k = 2.0 \times 10^{40} \, {\rm GeV^{-1}}$. The present CMB temperature is $T_{0} = 2.4 \times 10^{-13} \GeV$. With $g(T_{R\;max}) = 106.75$ for the Standard Model degrees of freedom and $g_{s}(T_{0}) = 3.91$, we obtain   
\be{e63}
N = \ln{\left[\frac{1.6 \times 10^{27}}{\alpha^{\frac{1}{4}}}\right]}  = 62.63-\frac{1}{4}\ln\left(\alpha \right)
~.\ee
This value of $N$ for a given $\alpha$ will be used later to calculate the value of $n_{s}$ for comparison with the observed results from Planck.

\section{Consistency with Unitarity Conservation}

In the Einstein frame, the Lagrangian term responsible for unitarity violation is 
\be{u1} \frac{\alpha}{4 M_{pl}^{4} } 
\frac{\left(\partial_{\mu} \phi \partial^{\mu} \phi \right)^{2}}{\left(1 + \frac{4 \alpha V(\phi)}{M_{pl}^4}\right)}  ~.\ee
In terms of the canonically normalised field $\sigma$, this becomes 
\be{u2}  \frac{\alpha}{4} 
\frac{\left(\partial_{\mu} \sigma \partial^{\mu} \sigma \right)^{2} }{M_{pl}^{4}} \left(1 + \frac{4 \alpha V(\phi)}{M_{pl}^4} \right)  ~.\ee
Expanding the rescaled canonically normalised inflaton field $\sigma$ about the classical inflaton background $\bar{\sigma}\left(t\right)$ 
\be{u3}
\sigma=\bar{\sigma}\left(t\right) + \delta \sigma
~,\ee
where $\delta \sigma$ describes the quantum perturbations around the classical background, we obtain the interaction term
\be{u3}  \frac{\alpha}{4} 
\frac{\left(\partial_{\mu} \delta \sigma \partial^{\mu} \delta \sigma \right)^{2}}{M_{pl}^{4}} \left(1 + \frac{4 \alpha V(\bar{\phi)}}{M_{pl}^4} \right)  ~,\ee
where $\bar{\phi} = \phi(\bar{\sigma})$.
The amplitude for $\delta \sigma \, \delta \sigma \rightarrow \delta \sigma \, \delta \sigma $ scattering
is therefore dimensionally given by
\be{u4} \mid \mathcal{M} \mid  \approx \frac{\alpha}{4} \frac{\tilde{E}^{4}}{M_{pl}^{4}}\left(1 + \frac{4 \alpha V\left(\bar{\phi}\right)}{M_{pl}^{4}}\right)
~,\ee
where $\tilde{E}$ is the energy calculated in the Einstein frame. 
Unitarity is violated for the scattering process once $ \mid \mathcal{M} \mid \gae 1$. This happens once $\tilde{E} \gae \tilde{\Lambda}$, where $\tilde{\Lambda}$ is the unitarity cutoff in the Einstein frame. Therefore 
\be{u6}
\tilde{\Lambda} \approx \frac{\sqrt{2} M_{pl}}{\alpha^{\frac{1}{4}}}\frac{1}{\left(1 + \frac{4\alpha V\left(\bar{\phi}\right)}{M_{pl}^{4}}\right)^{\frac{1}{4}}} ~.
\ee
In this expression, the value of $V(\bar{\phi})$ at $N$ e-foldings is given by $V(\bar{\phi}) = 2 m_{\phi}^{2} M_{pl}^{2} N $. Therefore, using $4 \alpha V(\bar{\phi})/M_{pl}^{4} \gg 1$, we obtain 
\be{u6a}  \tilde{\Lambda} \approx \frac{M_{pl}^{2}}{\alpha^{1/2} \left( 2 m_{\phi}^{2} M_{pl}^{2} N \right)^{1/4} }    ~.\ee 

The minimum condition that needs to be satisfied to keep unitarity violation in check is that the energy scale of the quantum fluctuations during inflation, which is approximately equal to the Hubble expansion rate calculated in the Einstein frame, $\tilde{H}$, should be less than the value of the cutoff scale during inflation, 
$ \tilde{H} < \tilde{\Lambda}$. With $\tilde{H} = M_{pl}/(12 \alpha)^{1/2}$, this requires that 
\be{u8} \left(\frac{m_{\phi}^{2}}{M_{pl}^{2}} \right) N \lae 72  ~.\ee
Since $m_{\phi}^{2}/M_{pl}^{2} \approx 3.4 \times 10^{-11}$, this is easily satisfied. Therefore the Palatini $\phi^{2} R^{2}$ inflation model is easily consistent with the condition for unitarity conservation during inflation. 

It is also interesting to calculate the unitarity violation scale in the present vacuum. This is given by setting $4 \alpha V/M_{pl}^{4} = 0$ in \eq{u6}, which gives  
\be{u9} \Lambda = \frac{\sqrt{2} M_{pl}}{\alpha^{1/4}}    ~.\ee 
For the model to be consistent with unitarity, either new physics must enter at a $\phi$ particle scattering energy below $\Lambda$, or the scattering must become non-perturbative but unitary at this energy.

\section{Results for the Case of Instantaneous Reheating}

In Figure 1 we show the value of $n_{s}$ as a function of $\alpha$, together with the 1-$\sigma$ and 2-$\sigma$ bounds from Planck.  In Table 1 we give the values of the scalar spectral index, $n_{s}$, the tensor-to-scalar ratio, $r$, the number of e-folds  corresponding to the Planck pivot scale, $N$, the instantaneous reheating temperature, $T_{R\;max}$, the inflaton at $N$ e-foldings, $\sigma(N)$, the Hubble parameter at the end of inflation, $\tilde{H}$,   
 the unitarity cutoff in the Einstein frame, $\tilde{\Lambda}$, and the unitarity cutoff in the present vacuum, $\Lambda$, for the different lower bounds on $\alpha$ derived in our discussion.

\begin{figure}[htbp]
\begin{center}
\epsfig{file=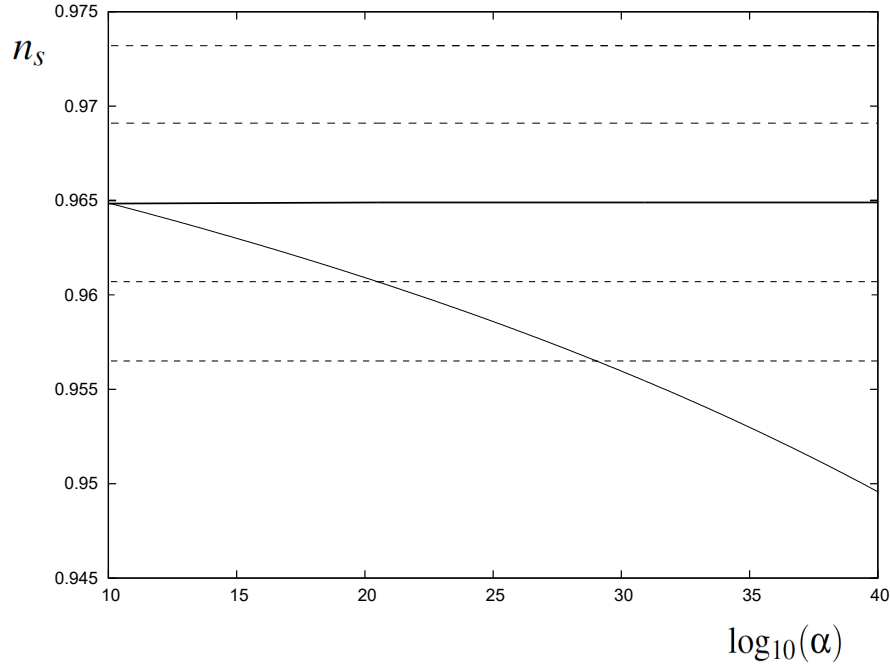, width=0.5\textwidth, angle = 90}
\caption{$n_{s}$ as a function of $\alpha$. The Planck best-fit and 1-$\sigma$ and 2-$\sigma$ bounds are also shown.} 
\label{fig2}
\end{center}
\end{figure}

\begin{table}[htbp]
\begin{center}
\begin{tabular}{ |c|c|c|c|c|c|c|c|c| }
\hline
$\alpha$ 
& $n_{s}$ 
& $r$ 
& $N$ 
& $T_{R\;max} /{\rm GeV}$
& $\sigma(N)/{\rm GeV}$  
& $\tilde{H} /{\rm GeV}$ 
& $\tilde{\Lambda} /{\rm GeV}$ 
& $\Lambda /{\rm GeV}$ 
\\
\hline
$10^{12}$ 
& $0.9641$ 
& $9.5 \times 10^{-6}$ 
& $55.7$
& $7.0 \times 10^{14}$  
& $1.5 \times 10^{18}$  
& $6.9 \times 10^{11}$ 
& $3.1 \times 10^{14}$ 
& $3.4 \times 10^{15}$ 
\\
\hline
$10^{20}$ 
& $0.9609$ 
& $1.1 \times 10^{-13}$ 
& $51.1$ 
& $7.0 \times 10^{12}$ 
& $4.3 \times 10^{14}$  
& $6.9 \times 10^{7}$ 
& $3.1 \times 10^{10}$ 
& $3.4 \times 10^{13}$ 
\\
\hline
$10^{32}$ 
& $0.9548$ 
& $1.5 \times 10^{-25}$ 
& $44.2$ 
& $7.0 \times 10^{9}$
& $8.0 \times 10^{8}$  
& $70$ 
& $3.3 \times 10^{4}$ 
& $3.4 \times 10^{10}$ 
\\
\hline
\end{tabular}
\caption{The scalar spectral index, $n_{s}$, the tensor-to-scalar ratio, $r$, the number of e-folds corresponding to the Planck pivot scale, $N$, the instantaneous reheating temperature, $T_{R\;max}$, 
$\sigma(N)$, the Hubble parameter at the end of inflation, $\tilde{H}$,   
the unitarity cutoff in the Einstein frame, $\tilde{\Lambda}$,  and the unitarity cutoff in the present vacuum, $\Lambda$, as a function of the lower bound on $\alpha$.}
\end{center}
\end{table}

We find agreement between the $n_{s}$ values and the 2018 Planck analysis for two out of three of the values of $\alpha$ considered. The value of the scalar spectral index and 1-$\sigma$ errors from the 2018 analysis  \cite{Planck}, assuming as priors $\Lambda$CDM and no running of the spectral index, is
\be{e64}
n_{s}= 0.9649 \pm 0.0042
\ee
with the 2-$\sigma$ lower bound given by $n_{s} > 0.9565$. It is clear that the values of the scalar spectral index for $\alpha = 10^{12}$ and $\alpha  = 10^{20}$ are easily within 1-$\sigma$ agreement with Planck. The case with $\alpha = 10^{32}$, corresponding to Planck suppressed corrections with no shift symmetry, is slightly below the 2-$\sigma$ lower bound from Planck for $\Lambda$CDM. 
We note that the status of the $\Lambda$CDM analysis is presently unclear due to the $H_{0}$ tension between local distance and Planck CMB determinations of  $H_{0}$. As a result, we can really only conclude that $\alpha = 10^{32}$ is likely  to be close to the 2-$\sigma$ lower bound on $n_{s}$\footnote{In \cite{martinelli}, the $H_{0}$ problem is addressed by including a 
time-dependent dark energy equation of state, which modifies the 1-$\sigma$ bound on $n_{s}$. For example, for the case of Planck CMB  $+$ Pantheon supernova data, the $n_{s}$ value is altered from  0.9653 $\pm$ 0.0046 to 0.9641 $\pm$ 0.0048, corresponding to a shift in the 2-$\sigma$ lower bound from 0.9561 to 0.9545. This would bring the $\alpha =  10^{32}$ result to within the 2-$\sigma$ range.}.

The post-inflation cosmology of the model is quite conventional, in spite of the large values of $\alpha$ considered, with reheating temperatures approximately in the range $10^{10} - 10^{15}$ GeV in the case of instantaneous reheating. We will consider some specific reheating mechanisms in the next section.  

In general, the tensor-to-scalar ratio is highly suppressed in models which have sub-Planckian values for $\sigma$, with $r \lae 10^{-5}$. This 
will be unobservable in the next generation of CMB experiments, which have a sensitivity $\delta r \sim 10^{-3}$.  

We have already noted that the value of $\tilde{H}$ during inflation is consistent with the minimal condition for unitarity conservation during inflation, $\tilde{H} < \tilde{\Lambda}$. A stronger condition for the model to be safe with respect to unitarity violation would be that the field $\sigma$ is less than $\tilde{\Lambda}$. However, this is not satisfied in these models. Therefore, either non-renormalisable potential corrections due to the new physics of unitarity conservation, scaled by $\tilde{\Lambda}$, would have to be suppressed, or unitarity conservation at high energies would have to be due to strong coupling in scattering processes at $\tilde{E} > \tilde{\Lambda}$, without the need for new physics and the associated potential corrections. 

It is interesting to consider the possible implications of $\alpha \approx 10^{32}$ being marginally excluded by CMB data. In this case, a small correction to the potential would be necessary to increase $n_{s}$ and bring the model into agreement with observation. If $\alpha$ is close to $10^{32}$, the Planck-suppressed corrections could themselves modify the predicted spectral index. In this interpretation of the tension between the model and observation,  the value of $\alpha$  would be fixed by the observed spectral index to be approximately $10^{32}$.  Alternatively, quantum corrections associated with the couplings of the inflaton to Standard Model particles, which are necessary for reheating, could modify the potential and so increase $n_{s}$. We will discuss this possibility further in the next section. The tension between the model and observation could also be resolved if the dimensionless coupling $k$ in the Planck-suppressed operator \eq{e32} were smaller than 1. For example, if $k \sim 0.001$ rather than $k \sim 1$ then the lower bound would become $\alpha \gae 10^{29}$ and the model would be within the 2-$\sigma$ lower bound. A smaller value of $k \sim 0.001$ is appropriate if we take the view that the interaction \eq{e32} should include a combinatorial factor $1/6!$, so that the coupling in the corresponding Feynman rule is of the order of $1/M_{pl}^{2}$.

\section{Reheating Mechanisms, Quantum Corrections and Condensate Stability} 

  So far we have considered the case of instantaneous reheating. The process of reheating to produce thermal Standard Model degrees of freedom will depend upon how the inflaton couples to the Standard Model. We will consider two natural couplings of a singlet inflaton to the Standard Model and its natural extension to include right-handed neutrinos:  a Higgs portal coupling and a coupling to right-handed neutrinos,  
\be{r1}  \frac{\lambda_{\phi H}}{2}  \phi^{2} |H|^{2} 
+ \left( \frac{\lambda_{\phi N}}{2} \phi \bar{N_{R}^{c}} N_{R} + \;h.c. \right) ~.\ee
These couplings will produce corrections to the inflaton potential, which will impose upper bounds on the couplings. We therefore first consider the 1-loop effective potential due to these couplings and the upper bound from the requirement that the 1-loop correction does not affect the prediction for the spectral index. 

\subsection{The 1-loop effective potential and $n_{s}$} 

We will calculate the 1-loop effective potential in the Jordan frame and then transfer the complete effective potential to the Einstein frame. The 1-loop effective potential in the Jordan frame is given by the Coleman-Weinberg expression\footnote{The $\alpha R^2/4$ term in the action could contribute terms proportional to $\alpha$ to the Jordan frame effective potential.  However, since these $\alpha$-dependent terms are in addition to the Coleman-Weinberg potential, they will not affect the requirement that the Jordan frame Coleman-Weinberg potential should not perturb the predictions of the model.}   
\be{c1} \Delta V_{CW}(\phi) = \sum_{i} \pm \frac{M_{i}^{4}(\phi)}{64 \pi^{2}} \ln \left( \frac{M_{i}(\phi)^{2}}{\mu^{2}} \right)   ~,\ee   
where the sum is over bosonic and fermionic degrees of freedom, with a $+$ $(-)$ sign for bosons (fermions). 
The complete Jordan frame effective potential is then 
\be{c2} V_{TOT}(\phi) = V(\phi) + \Delta V_{CW}(\phi) ~.\ee
Since in this case the potential corrections $\Delta V_{CW}$ are defined in the Jordan frame, we can use the equivalence, demonstrated in \cite{Enckell}, of the spectral index calculated for the conventional chaotic inflation model with potential $V_{TOT}(\phi)$ to that of the Palatini $R^2$ model based on $V_{TOT}(\phi)$. The spectral index is then 
\be{c1} n_{s} = 1 + 2 \overline{\eta} - 6\overline{\epsilon}   ~,\ee
where 
\be{c2}  \overline{\eta} = M_{pl}^{2} \frac{\partial^{2} V_{TOT}}{\partial \phi^{2}} \;\;\;;\;\;\; \overline{\epsilon}=\frac{M_{pl}^{2}}{2} 
\left( \frac{\partial V_{TOT}}{\partial \phi}\right)^{2}  ~, \ee
and $\phi(N) = 2 \sqrt{N} M_{pl}$. 
Keeping terms to leading order in $\Delta V_{CW}$, the shift of the spectral index due to the 1-loop correction is given by
\be{c3} \Delta n_{s} = 2 \overline{\eta} \left( \frac{\Delta V_{CW}''}{V''} - \frac{\Delta V_{CW}}{V} \right) - 6 \overline{\epsilon} \left( \frac{2 \Delta V_{CW}'}{V'} - \frac{2 \Delta V_{CW}}{V} \right)  ~.\ee 
The resulting corrections to $n_{s}$ from $\lambda_{\phi H}$ and $\lambda_{\phi N}$ are then 
\be{c4a} \Delta n_{s\;H} = \frac{M_{pl}^{2}}{2 \pi^{2}} \frac{\lambda_{\phi H}^{2}}{m_{\phi}^{2}} \left[ 1 - \ln\left( \frac{\lambda_{\phi H} \phi^{2}}{\mu^{2}} \right) \right]   ~\ee
and 
\be{c4b} \Delta n_{s\;N} = - \frac{M_{pl}^{2}}{4 \pi^{2}} \frac{\lambda_{\phi N}^{4}}{m_{\phi}^{2}} \left[ 1 - \ln\left( \frac{\lambda_{\phi N}^{2} \phi^{2}}{\mu^{2}} \right) \right]     ~.\ee
We will choose the renormalisation scale $\mu$ such that the 
logarithmic term in the correction is zero when $\phi = \phi(N)$, where $N$ corresponds to the Planck pivot scale, with $m_{\phi}$ then defined at this scale. We should also include a tree-level $\lambda_{\phi} \phi^4$ term in the renormalisable potential, so we are assuming that this is zero or negligible at the renormalisation scale in order to be consistent with the $\phi^2$ classical potential upon which the model is based. The corrections are then
\be{c5a} \Delta n_{s\;H} = \frac{\lambda_{\phi H}^{2}}{2 \pi^{2}} \frac{M_{pl}^{2}}{m_{\phi}^{2}} ~\ee
and 
\be{c5b}  \Delta n_{s\;N} = - \frac{\lambda_{\phi N}^{4}}{4 \pi^{2}} \frac{M_{pl}^{2}}{m_{\phi}^{2}}  ~. \ee
Requiring that $|\Delta n_{s}| < 0.001$, we obtain the upper bounds 
\be{e6a} \lambda_{\phi H} < \left(\frac{0.001 \times 2 \pi^{2} m_{\phi}^{2}}{ M_{pl}^{2}}\right)^{1/2} = 8.2 \times 10^{-7}  ~\ee
and
\be{e6b}   \lambda_{\phi N} < \left(\frac{0.001 \times 4 \pi^{2} m_{\phi}^{2}}{ M_{pl}^{2}}\right)^{1/4} = 1.1 \times 10^{-3}  ~.\ee 
We note that the quantum correction to the potential could increase the spectral index and so allow values of $\alpha$ greater than $10^{32}$ to be in agreement with the 2-$\sigma$ Planck lower bound on $n_{s}$. For this to happen, $\Delta n_{s}$ must be positive, which is true for the correction due to the Higgs portal coupling.

\subsection{Reheating via decay to right-handed neutrinos}

Since we will be considering a condensate in the regime $\phi < \phi_{0}$, the Jordan and Einstein frames become equivalent, with $\sigma \approx \phi$. Therefore we will discuss reheating in terms of $\phi$ when $\phi < \phi_{0}$.

Assuming that the right-handed neutrino mass is small compared to $m_{\phi}$, the decay rate of the $\phi$ scalars to right-handed neutrinos is given by
\be{r01} \Gamma_{\phi \rightarrow NN} = \frac{\lambda_{\phi N}^{2} m_{\phi}}{16 \pi}   ~.\ee
The condition for instantaneous reheating is that 
\be{r02} \Gamma_{\phi \rightarrow NN} > \tilde{H} = \frac{M_{pl}}{\sqrt{12 \alpha}}   ~.\ee
This is satisfied if 
\be{r03} \lambda_{\phi N} > \left( \frac{64 \pi^{2} M_{pl}^{2}}{3 \alpha m_{\phi}^{2} } \right)^{1/4}  \equiv 1.6\times 10^{-5} \times \left(\frac{10^{32}}{\alpha} \right)^{1/4}  ~.\ee
The upper bound on $\lambda_{\phi N}$ from the correction to $n_{s}$ is  $\lambda_{\phi N} < 1.1 \times 10^{-3}$, 
so for the case of general Planck-scale suppressed corrections to the potential, which require $\alpha \gae 10^{32}$, it is generally possible to have a large enough $\lambda_{\phi N}$ to have instantaneous reheating without introducing too large quantum corrections into the potential. For values of $\alpha$ much smaller than $10^{32}$, such as the limit at which Planck corrections with a shift symmetry are suppressed, $\alpha \approx 10^{20}$, reheating by this mechanism cannot be instantaneous and the model will therefore have a lower reheating temperature than previously estimated. This will cause $n_{s}$ to be lower. However, models with $\alpha$ much smaller than $10^{32}$ have values of $n_{s}$ that are not close to the 2-$\sigma$ lower bound and so can undergo reheating at a lower temperature whilst remaining consistent with the observed $n_{s}$.

\subsection{Reheating via the Higgs portal}  

In the case of reheating via the Higgs portal, the process is annihilation of the inflaton condensate scalars to Higgs bosons. The case where the annihilation is Bose-enhanced corresponds to preheating \cite{preheating}, with the creation of relativistic Higgs bosons in a momentum state with $k = m_{\phi}$. However, whether preheating can occur will depend upon whether the condensate undergoes fragmentation. 

If the condensate fragments, the $\phi$ scalars are bound together in oscillons of diameter $\sim m_{\phi}^{-1}$. This has two consequences for annihilation of the $\phi$ scalars in the condensate. First, the number density of $\phi$ scalars in the oscillons does not decrease, unlike the case of $\phi$ scalars in a coherent condensate where $n_{\phi} \propto 1/a^3$. As a result \cite{jmf}, the annihilation rate of the $\phi$ scalars in the oscillons is {\it constant}, therefore $\Gamma_{ann} > \tilde{H}$ will eventually be satisfied and reheating via annihilation can occur. 
This is in contrast to the case of annihilation of 
scalars in a coherently oscillating condensate, in which case $ \Gamma_{ann} \propto n_{\phi} \propto 1/a^{3}$, compared to $\tilde{H} \propto \left(\rho_{\phi}\right)^{1/2} \propto 1/a^{3/2}$, and so $\Gamma_{ann}$ decreases faster with expansion that $\tilde{H}$. Therefore, unless annihilation is fast enough to reheat immediately when the $\phi$ condensate first forms, reheating via annihilation is not possible for a coherent condensate. 
 
A second consequence of fragmentation is that preheating is unlikely occur. This is because relativistic 
Higgs particles created via annihilation of the zero-momentum scalars in the condensate will rapidly escape from the volume of the oscillon and therefore the relativistic momentum mode of the scalar field cannot become occupied inside the oscillon. Thus no Bose enhancement of the annihilation process can occur. 

Therefore, to estimate the reheating temperature via annihilation, we must first check if the condensate fragments and, if so, we should compute the perturbative annihilation rate of the scalars in the oscillons. 

We have shown that inflation ends and rapid rolling of the  inflaton $\sigma$ begins when the inflaton is on the plateau of the potential. In this case it is highly likely that that fragmentation will rapidly occur via tachyonic preheating \cite{tp}. To check this, we will use an analytical condition derived in \cite{jk1}. This considers whether fragmentation will occur after a coherent condensate has formed, even if tachyonic preheating does not cause even faster fragmentation. As such, it provides a sufficient condition for fragmentation.

For a potential of the form 
\be{r3} V = \frac{1}{2} m_{\phi}^{2} \phi^{2} - A \phi^{4}   ~,\ee
where it is assumed that the potential is dominated by the quadratic term, 
the condition for fragmentation is that \cite{jk1}
\be{r4}  A > \frac{100}{r_{V}} \frac{m_{\phi}^{2}}{M_{Pl}^{2}}   ~.\ee 
Here $r_{V} < 1$ is the ratio of the quartic to the quadratic part of the potential when the oscillations begin, which we can choose to be $r_{V} = 0.1$. 

At $\phi < \phi_{0}$, the inflaton potential of the  Palatini $\phi^2$ $R^2$ model has the form 
\be{r5} V_{E}(\phi) = \frac{V(\phi)}{1 + \frac{4 \alpha V(\phi)}{M_{Pl}^{4}} } \approx V(\phi)\left(1 - \frac{4 \alpha V(\phi)}{M_{Pl}^{4}} \right)   ~.\ee
Thus 
\be{r6}  V \approx \frac{1}{2} m_{\phi}^{2}\phi^{2} - \frac{\alpha m_{\phi}^{4}}{M_{Pl}^{4}} \phi^{4}  ~.\ee 
Therefore $A = \alpha m_{\phi}^{4}/M_{Pl}^{4}$. A sufficient condition for fragmentation is then             
\be{r7} \frac{\alpha m_{\phi}^{2}}{M_{pl}^{2}}  > 1000 ~.\ee 
This is very strongly satisfied for the cases with Planck-suppressed potential corrections, where $\alpha m_{\phi}^{2}/ M_{pl}^{2} \approx 10^{9}$ for $\alpha = 10^{20}$  and $\alpha m_{\phi}^{2}/M_{pl}^{2} \approx 10^{21}$ for $\alpha = 10^{32}$. So in these cases we can expect almost instantaneous fragmentation to occur. The condition for fragmentation is not satisfied for the limiting case of a sub-Planckian $\phi$, $\alpha \approx 10^{12}$, for which $\alpha m_{\phi}^{2}/ M_{pl}^{2} \approx 30$. 
However, the condition \eq{r4} underestimates the formation of oscillons, therefore it is still possible that oscillons will form in this case.

Assuming that fragmentation occurs, reheating will occur via perturbative annihilation of zero-momentum scalars in the oscillons. For rapid fragmentation, the energy density in the oscillons will approximately equal the energy density of the inflaton at the end of inflation 
\be{r8}  \rho \approx  3 \tilde{H}^{2} M_{Pl}^{2} \approx \frac{M_{Pl}^{4}}{4 \alpha}  ~.\ee
Therefore the $\phi$ number density in the fragments is 
\be{r9} n_{\phi} = \frac{\rho_{\phi}}{m_{\phi}} = \frac{M_{Pl}^{4}}{4 \alpha m_{\phi} } ~.\ee
The annihilation cross-section times relative velocity for $\phi \phi \rightarrow h_{i} h_{i}$ ($i = 1,...  ,4$), where we can consider the four real scalars in the Higgs doublet to be physical, is
\be{r10}  <\sigma_{ann} v> = \frac{\lambda_{\phi H}^{2}}{16 \pi m_{\phi}^{2} } ~.\ee
So the perturbative annihilation rate is 
\be{r11} \Gamma_{ann} = n_{\phi} <\sigma_{ann} v> = \frac{\lambda_{\phi H}^{2} M_{pl}^{4}}{64 \pi \alpha m_{\phi}^{3} }  ~.\ee
The condition for reheating via annihilation of the scalars in the oscillons is 
\be{r12} \Gamma_{ann} = \tilde{H} \equiv \frac{k_{T} T_{R}^{2}}{M_{Pl}} \;\;;\;\; k_{T} =\left(
\frac{\pi^{2} g(T)}{90}\right)^{1/2} ~.\ee 
Therefore, using $k_{T} = 3.3$ for $g(T) \approx 100$,
\be{r13} T_{R} = \frac{ \lambda_{\phi H}}{ 
\left(64 \pi \alpha m_{\phi}^{3}  \right)^{1/2}} \frac{M_{Pl}^{5/2}}{k_{T}^{1/2}}  
 \approx  660 \GeV \times \left(\frac{\lambda_{\phi H}}{10^{-6}}\right) \left(\frac{10^{32}}{\alpha}\right)^{1/2}   ~.\ee
Using the upper bound on $\lambda_{\phi H}$ from $\Delta n_{s\;H}$, $\lambda_{\phi H} < 8.2 \times 10^{-7}$,  
we find   
\be{r15a}   T_{R} \lae  500   \left(\frac{10^{32}}{\alpha}\right)^{1/2}   \GeV   ~.\ee
Thus for $\alpha = 10^{20}$ we have $T_{R} \lae 5 \times 10^{8} \GeV$, whilst for $\alpha = 10^{32}$ we have $T_{R} \lae 500 $ GeV. Thus in both cases successful reheating can be achieved, but the reheating temperatures are well below the corresponding instantaneous reheating temperatures, $T_{R\;max}$, given in Table 1. Since lower $T_{R}$ will result in a lower value for $n_{s}$, this suggests that for the case $\alpha \gae 10^{32}$,
reheating via the Higgs portal coupling is disfavoured by the observed value of $n_{s}$. This estimate for $T_{R}$ is based on a number of simplifying assumptions, in particular the stability of the oscillon throughout inflaton annihilation. However, this assumption favours the annihilation process and so will lead to a maximum possible reheating temperature via annihilation to Higgs bosons\footnote{ 
We note that the term in the action quartic in the derivative of $\sigma$ may become significant after slow-roll inflation and during reheating, which could modify the conventional analysis of fragmentation via tachyonic preheating. We thank Antonio Racioppi for bringing this to our attention.}.

\section{Conclusions}

The idea of inflation based on a minimal $\phi^{2}$ potential, corresponding to the simplest potential for a massive scalar field, has an appealing simplicity and is interesting from a particle physics model-building point of view. In this work we have shown that, in addition to solving the large $r$ problem of the original $\phi^2$ chaotic inflation model, the Palatini $\phi^2 R^2$ model of \cite{Enckell} and \cite{Antoniadis} can also solve the super-Planckian inflaton problem of $\phi^2$ chaotic inflation and can be consistent with Planck-suppressed potential corrections such as could arise from a quantum gravity completion. In addition, we have shown that the model reheats to a temperature sufficient for nucleosynthesis and conserves unitarity during inflation. 
As such, the Palatini $\phi^2$ $R^2$ model provides a completely consistent inflation model with a viable post-inflation cosmology.

We have determined the lower bounds on the dimensionless parameter of the $R^2$ term, $\alpha$, for which the Palatini $\phi^{2} R^2$  model can be consistent with a sub-Planckian inflaton and with Planck-suppressed potential corrections, both in the case of general Planck-suppressed corrections and in the case of corrections with a broken shift symmetry. We find that $\alpha \gae 10^{12}$ is necessary to have a sub-Planckian inflaton, $\alpha \gae 10^{20}$ is necessary for the scalar spectral index to be unaffected by Planck-suppressed corrections with a broken shift symmetry, and $\alpha \gae 10^{32}$ is necessary for general Planck-suppressed corrections.

The values of $\alpha$ in the sub-Planckian Palatini $\phi^2 R^2$ model are larger than those in the conventional Starobinsky $R + R^2$ model \cite{starobinsky}, which also requires a very large dimensionless coupling, $\alpha \approx 10^{10}$. In general, without a metric by which to gauge the significance of the very large dimensionless couplings that are a common feature of inflation models based on non-minimal and higher-order gravitational interactions, there is no a priori reason to disfavour such models.

We have calculated the reheating temperature in the case of instantaneous reheating, corresponding to the maximum possible reheating temperature. We find that as the value of $\alpha$ increases, the reheating temperature decreases, with $T_{R\;max} \sim 10^{10}$ GeV for the case of $\alpha = 10^{32}$, corresponding to the lower bound on $\alpha$ from general Planck-suppressed potential corrections. Therefore the model can reheat to a high enough temperature for a viable post-inflation cosmology.

Given the reheating temperature, we can determine the number of e-foldings corresponding to the Planck pivot scale and hence check that the predicted spectral index is in agreement with observations. 
For the case of instantaneous reheating, we find that the predicted values of the scalar spectral index are in agreement with the most recent Planck results \cite{Planck} to within 1-$\sigma$ for $\alpha \approx 10^{12}$ and $\alpha \approx 10^{20}$. For the case of general Planck-suppressed corrections, we find that $\alpha \approx 10^{32}$ is close to the 2-$\sigma$ lower bound. This may indicate that the model favours the limit with $\alpha \approx 10^{32}$, where the Planck-suppressed potential corrections are large enough to modify the predictions of the model and so increase $n_{s}$ sufficiently to bring it into better agreement with the value from Planck. Alternatively, quantum corrections due to the couplings responsible for reheating could increase the $n_{s}$ prediction. The tension with observation could also be resolved if the dimensionless constant $k$ in the Planck-suppressed operator were less than $k \sim 1$. For example, $k \sim 0.001$ due to a combinatorial factor would reduce the lower bound to $\alpha \gae 10^{29}$ and so could allow the model to be within the 2-$\sigma$ lower bound.  

As $\alpha$ increases, the tensor-to-scalar ratio $r$ becomes increasingly suppressed. For the case of sub-Planckian inflation, $r$ is less than $10^{-5}$ and therefore unobservable in the next generation of CMB experiments. 

The model introduces interactions which violate perturbative unitarity. We find that the unitarity violation scale during inflation in the Einstein frame, $\tilde{\Lambda}$, is generally much larger than the expansion rate $\tilde{H}$. Therefore the model is generally consistent with the minimal condition for unitarity conservation during inflation,  $\tilde{H} 
< \tilde{\Lambda}$. However, it should be noted that the unitarity violation scale is smaller than the inflaton field during inflation. Therefore non-renormalisable potential corrections associated with a unitarity-conserving completion would exclude the model. This can be avoided if the perturbative unitarity violation scale is in fact a strong coupling scale, with unitarity being conserved non-perturbatively \cite{HIuv}. However, as in the case of Higgs inflation \cite{HIuv2,HIuv3}, unitarity conservation, whilst not excluding the model, is not a trivial issue for this model.

We considered two specific reheating mechanisms: reheating via inflaton annihilation to Higgs bosons and reheating via inflaton decay to right-handed neutrinos. 
We find that the inflaton condensate is likely to fragment, resulting in oscillon formation. After placing an upper bound on the Higgs portal coupling from quantum corrections to the potential, we estimated the reheating temperature from inflaton annihilation to Higgs bosons in oscillons and found that reheating is not instantaneous and could result in a low reheating temperature, less than 500 GeV for the case with $\alpha \gae 10^{32}$. This would result in a value for the spectral index well below the 2-$\sigma$ observational lower bound. In contrast, if the Standard Model is extended to include right-handed neutrinos, inflaton decay to right-handed neutrinos can be rapid enough to produce instantaneous reheating and a high reheating temperature. 

The Palatini $\phi^{2}$ $R^{2}$ inflation model is an interesting addition to the class of minimal inflation models. We have shown that, in addition to being consistent with Planck observations, the model can also be consistent with sub-Planckian inflation and with potential corrections from quantum gravity, whilst conserving unitarity during inflation and reheating to a sufficient temperature for a successful post-inflation cosmology.

\section*{Note Added}

While this work was in progress, a paper that also considers the Palatini $\phi^2 R^2$ inflation model appeared on arXiv \cite{RecentTommi}. This also considers very large values of $\alpha \gae 10^{37}$, with a quite different motivation from that considered here.  We find that where our results overlap with those of \cite{RecentTommi}, they are broadly in agreement.  In addition, a paper discussing reheating in Palatini $R^2$ models has recently been appeared on arXiv \cite{reheating}, which generalises the analysis of $n_{s}$ to beyond the case of instantaneous reheating.

\section*{Acknowledgements}

We thank Antonio Racioppi for his comments.   The work of ALS is supported by STFC.

\end{document}